\documentstyle[12pt,preprint,tighten,floats,aps,epsf,psfig]{revtex}
\def\simlt{\stackrel{<}{{}_\sim}}
\def\simgt{\stackrel{>}{{}_\sim}}
\preprint{IC/98/186}
\tighten
\begin{document}
\draft
\title{Weak-scale hidden sector and electroweak Q-balls} 
\author{D. A. Demir}
\address{The Abdus Salam International Centre for Theoretical Physics,
I-34100 Trieste, Italy}
\date{\today}
\maketitle
\begin{abstract}
By extending the scalar sector of the Standard Model (SM) by a $U(1)$ singlet, 
we show
that the electroweak symmetry breaking enables the formation of a stable,
electrically neutral, colorless Q-ball which couples to the SM particle spectrum
solely through the Higgs boson. This Q-ball has mainly weak and gravitational
interactions, and behaves as a collection of weakly interacting massive particles.
Therefore, it can be a candidate for the dark matter in the universe.\\
\end{abstract}
\newpage
Classification of the total particle spectrum as those residing in the
'hidden sector' and those of the 'observable sector' has played a crucial 
role in high-scale supersymmetric theories for the implementation of
the supersymmetry breaking at low energies. Indeed, the soft supersymmetry 
breaking terms in the low-energy globally supersymmetric Lagrangian are 
mediated by either gravity \cite{sugra} (supergravity models) or messenger fields
\cite{gmsb} (gauge-mediated supersymmetry breaking models) from a hidden sector 
containing heavy singlets. Similar ideas, with essential differences 
from the supersymmetric case, also have been proposed for the scalar 
sector of the Standard Model (SM) \cite{old-hiding}. In the latter case one extends
the scalar sector by a gauge singlet which interacts only with the SM Higgs 
doublet and, of course, gravity. Unlike the supersymmetric theories, in such 
models singlet field and the SM particle spectrum are not necessarily at 
diversely different mass scales; one can take both observable and hidden 
sectors around the same energy scale, that is, weak scale. The other important
difference lies in the fact that the hidden sector in the SM Lagrangian 
is designed to account for the non-observation of the Higgs particle at 
the colliders as detailed in \cite{new-hiding}. To utilize the large $N$
expansion technique, and to increase the available invisible decay channels for the
Higgs boson, in \cite{old-hiding,new-hiding} use has been made of an $O(N)$
symmetric SM singlet. However, as long as one is not interested in the calculation
of specific scattering processes involving the Higgs boson and the singlet the
requirement of an $O(N)$ singlet can be relaxed as we shall do below. 

Inspired from the idea of a hidden Higgs sector \cite{old-hiding,new-hiding}, below
we extend the SM Lagrangian by a complex SM singlet scalar which interacts only with
the usual Higgs doublet. Namely we extend the SM gauge group by an extra global
Abelian group factor, $U(1)_{s}$, under which none of the SM particle spectrum is
charged. Below we shall show that this is the simplest extension of the SM Higgs sector 
which accomodates the non-topological solitons
\cite{td-lee,freidberg}, or in particular,  Coleman's Q-balls \cite{s-coleman}.
Non-topological solitons are extended objects with finite mass and spatial
extension, and arise in scalar field theories when there is an exact continuous
symmetry and some kind of attractive interaction, as classified by Coleman 
\cite{s-coleman}. In what follows we show that electroweak symmetry breaking
produces the necessary interactions between the physical Higgs field and the singlet
so that electrically neutral, colorless non-topological solitons naturally appear
in the true electroweak vacuum.
 
After analyzing the properties of this extended scalar sector with subsequent
discussion of the Q-balls it will be seen that: (1) In the electroweak vacuum the
physical Higgs field and the singlet combine to form an absolutely stable Q-ball
whose
interactions with the SM particle spectrum is provided solely by the Higgs field.
(2) Due to the extension of the scalar sector, Lagrangian necessarily obtains extra
unknown parameters which, however, remain embedded in the expressions for the
extensive parameters of the Q-ball, and do not have any direct effect on the
interaction between the SM particle spectrum and the Q-ball. (3) Q-ball forms the
state
of minimal energy in the scalar sector of the SM, and thus, any scalar produced in
the true vacuum via some collision process immediately escapes to the Q-ball
implying that the collider search for the Higgs particle may not obtain a
significant signal. (4) Photon, gluon and light fermions, due to their loop
suppressed couplings to the Higgs particle, interact very weakly with the Q-ball
compared to the massive electroweak bosons and heavy fermions. In this sense the
Q-ball behaves as a stable collection of weakly interacting massive particles, and
thus forms a candidate for the dark matter in the universe. 

Below we work out the extended SM Higgs sector and show that the
above-mentioned points do naturally follow. The model Lagrangian is invariant under
the SM gauge group $G_{SM}=SU(3)_{c}\times SU(2)\times U(1)_{Y}$ and an extra
global Abelian group $U(1)_{s}$. Suppressing the contributions of the gauge bosons
and fermions, the Lagrangian reads 
\begin{eqnarray}
{\cal{L}}=\frac{1}{2}\partial_{\mu}S^{*}\partial^{\mu}S+
(\partial_{\mu}\Phi)^{\dagger}(\partial^{\mu}\Phi)-V(|S|,|\Phi|)\;.
\end{eqnarray}
Here the potential $V(|S|,|\Phi|)$ is defined by
\begin{eqnarray}
V(|S|,|\Phi|)=\frac{1}{2}m_{s}^{2}|S|^{2}+\frac{1}{4}\lambda_{s}|S|^{4}
+m^{2}|\Phi|^{2}+\lambda |\Phi|^{2}-\kappa |S|^{2}|\Phi|^{2}\; .
\end{eqnarray}
where $\Phi$ is the SM Higgs doublet and $S$ is the SM- singlet. As the form of the
potential suggests enlargement of the SM spectrum by a singlet brings about three
new parameters $m_{s}^{2}$, $\lambda_{s}$ and $\kappa$. For the potential to be
bounded
from below in $S$ and $\Phi$ directions it is necessary to have $\lambda_{s}>0$,
$\lambda>0$, and $\sqrt{\lambda\lambda_{s}}-\kappa>0$. In addition to these, for
$U(1)_{s}$ symmetry to remain unbroken one
needs $m_{s}^{2}>0$. The sign of $\kappa$ is irrelevant for calculating the Higgs
decay width in the true vacuum of the theory, however, for the formation of Q-balls
it should be positive. In passing we note that for $\kappa\sim
\lambda\sim\lambda_{s}$ 
contribution of the quartic terms resembles that of the D-terms in supersymmetric
theories. 

As usual, when $m^{2}$ is negative, $\Phi$ develops a non-zero vacuum expectation
value  $<\Phi>=(v/\sqrt{2},0)$ where $v=\sqrt{-m^{2}/\lambda}=246\; \mbox{GeV}$.
Around this vacuum expectation value $\Phi$  can be expanded as
$\Phi=(1/\sqrt{2})(v+h+i\pi_{Z}, \pi_{W}^{1}+i\pi_{W}^{2})$ where $\pi_{Z}$
($\pi_{W}^{1,2}$) are the Golstone bosons swallowed by $Z$ ($W^{\pm}$) to
acquire a mass. Therefore, in this minimum of the potential the SM gauge symmetry is
broken and the Lagrangian (1) takes the from
\begin{eqnarray}
{\cal{L}}=\frac{1}{2}\partial_{\mu}S^{*}\partial^{\mu}S+
\frac{1}{2}\partial_{\mu}h \partial^{\mu}h-V(|S|,h)
\end{eqnarray}
with
\begin{eqnarray}
V(|S|,h)=\frac{1}{2}(m_{s}^{2}-\kappa v^{2})|S|^{2}+\frac{1}{4}\lambda_{s}|S|^{4}
+\frac{1}{2}m_{h}^{2}h^{2}+\lambda v h^{3} +\frac{\lambda}{4}h^{4}-\kappa v h
|S|^{2}-\frac{1}{2}\kappa h^{2}|S|^{2}\;,
\end{eqnarray}
where $h$ is the physical Higgs boson with mass $m_{h}=\sqrt{-2 m^{2}}$ as usual,
and we subtracted the vacuum energy to make $V(0,0)=0$. As the form of $V(|S|,h)$
suggests, electroweak symmetry breaking modifies the mass of the singlet as
$m_{s}^{2}\rightarrow m_{s}^{2}-\kappa v^{2}$, which must be positive to leave
$U(1)_{s}$ unbroken. This can always be satisfied by choosing $m_{s}^{2}$  
large enough.

There are some important points deserving discussion about the Lagrangian (3).
Firstly, passage from (1) to (3) involves the electroweak symmetry breaking
$G_{SM}\rightarrow SU(3)_{c}\times U(1)_{EM}$ so that experimentally
well-established SM particle spectrum arises. During this transition neither the
Higgs field, $h$, nor the other particles are affected by the presence of the
singlet field, that is, electroweak phase transition proceeds independently of the
interactions with the singlet. Secondly, the last two terms in (4) dictate the decay
and scattering of the Higgs particle to invisible matter. Indeed these operators
realize the processes $h \rightarrow S S$ and $h h \rightarrow S S$ whose signatures
are, of course, outside the experimental detection. It is the $h \rightarrow S S$
decay \cite{new-hiding} that is stressed in the literature to account for the
non-observation of the Higgs particle at the colliders. Finally, both Lagrangians
(1) and (3) are invariant under $U(1)_{s}$ so that $U(1)_{s}$ is a global
continious symmetry of the theory before and after the electroweak phase transition. 
   
There is a large amount of literature on non-topological solitons (see \cite{td-lee}
and references therein). Altough there are various types of such solitonic solutions
\cite{td-lee,freidberg} here we are interested in non-topological solitons of Q-ball
type
\cite{s-coleman}. In recent years interest in the physical implications of Q-balls
has accelerated after observing that they generically exist in supersymmetric
theories \cite{mssm-Q,enqvist1}. In general, Q-balls exist in scalar field theories
having a
continuous symmetry and some kind of attractive interaction among the scalars. They
are absolutely stable as long as the symmetry group is exact. Q-balls can arise from
the self interactions of a single scalar field \cite{s-coleman} as well as from the
interactions  among various scalar fields with different flavour \cite{mssm-Q}. In
fact, demonstration of the existence of Q-ball type solutions for the Lagrangian (3)
is nothing but a special case of the corresponding analyses in the supersymmetric
models \cite{mssm-Q,enqvist1} which contain Higgs fields and scalar fermions. Other
closely related works are \cite{kolb,freidberg} in which Q-ball formation in a model
containing one real
and one complex field was investigated. Although the Lagrangians employed in these
works are similar to (3) the real scalar field there has nothing to do with the
Higgs boson of the SM. Guided by the existing literature, below we present a short
description of
the Q-ball formation in the electroweak Lagrangian (3) referring the reader to
\cite{s-coleman,freidberg,mssm-Q,kolb} for details. $U(1)_{s}$, being an exact
symmetry of the
Lagrangians (1) and (3), has the conserved charge 
\begin{eqnarray}
Q=\frac{1}{2i}\int d^{3}\vec{x}S^{*}(\vec{x},t)
\stackrel{\leftrightarrow}{\partial}_{t} S(\vec{x},t)\;.
\end{eqnarray} 
Since $Q$ vanishes identically for the trivial solution $S(\vec{x},t)\equiv 0$, the
field configuration that minimizes the energy 
\begin{eqnarray}
E=\int d^{3} \vec{x}\{\frac{1}{2}|\partial_{t} S|^{2}+\frac{1}{2}|\nabla S|^{2}  
+\frac{1}{2}(\partial_{t} h)^{2}+\frac{1}{2}(\nabla h)^{2}+V(|S|,h)\}
\end{eqnarray} 
for $Q\neq 0$, must have non-vanishing values in a finite domain. This constrained
minimization can be accomplished by introducing a Lagrange multiplier $\omega$ and
minimizing 
\begin{eqnarray}
{\cal{E}}=E+\omega(Q-\frac{1}{2i}\int d^{3}\vec{x}S^{*}(\vec{x},t)
\stackrel{\leftrightarrow}{\partial}_{t} S(\vec{x},t))
\end{eqnarray}
with respect to fields and $\omega$, independently. A close inspection of
${\cal{E}}$ shows that those terms having an explicit time dependence can be
eliminated by requiring fields to rotate in the internal $U(1)_{s}$ space with
angular velocities proportional to their $U(1)_{s}$ charges \cite{s-coleman}, that
is,
$S(\vec{x},t)=e^{i\omega t}\bar{S}(\vec{x})$ and $h(\vec{x},t)=\bar{h}(\vec{x})$.
Here $\bar{S}$ and $\bar{h}$ are real and time-independent fields. With these
redefinitions ${\cal{E}}$ becomes
\begin{eqnarray}
{\cal{E}}=\int d^{3} \vec{x}\{\frac{1}{2}|\nabla \bar{S}|^{2}+
\frac{1}{2}(\nabla \bar{h})^{2}+V_{\omega}(\bar{S},\bar{h})\}+\omega Q
\end{eqnarray} 
where the effective potential $V_{\omega}$ is given by
\begin{eqnarray}
V_{\omega}(\bar{S},\bar{h})=V(\bar{S},\bar{h})-\frac{1}{2}\omega^{2}\bar{S}^{2}\; .
\end{eqnarray}
Consequently, the requirement of a finite $U(1)_{s}$ charge leads one to a new
potential $V_{\omega}$ which, unlike the original potential $V$ which has its global
minimum at ($S=0,h=0$), can develop a global minimum at some field configuration
away from the origin because of the $1/2 \omega^{2} \bar{S}^{2}$ term. In the true
electroweak vacuum there is a perturbative particle spectrum consisting of, in
addition to the usual SM spectrum, scalar bosons $S$ with unit $U(1)_{s}$ charge
and mass $\mu(0,0)=(m_{s}^{2}-\kappa v^{2})^{1/2}$, where
\begin{eqnarray}
\mu^{2}(\bar{S},\bar{h})\equiv \frac{2 V(\bar{S},\bar{h})}{\bar{S}^{2}} \;.
\end{eqnarray}
It is in this perturbative scheme that invisible decay rate of the Higgs particle
has been computed \cite{old-hiding,new-hiding}. However, the minimization of the
energy for a finite $U(1)_{s}$ charge modifies the original potential as in
$V_{\omega}$ and gives rise to appearence of new particles in the spectrum
\cite{s-coleman}. Indeed, if $\mu^{2}$ is minimized for some field configuration
$(\bar{S}=S_{0}\neq0, \bar{h}=h_{0}\neq 0)$, that is,
\begin{eqnarray}
\omega_{0}^{2}\equiv \frac{2
V(S_{0},h_{0})}{S_{0}^{2}}=Min[\mu^{2}(\bar{S},\bar{h})]<\mu^{2}(0,0)
\end{eqnarray}
then there exists non-dispersive solutions of the field equations ($\delta
{\cal{E}}/\delta \bar{S}=0\;$, $\delta {\cal{E}}/\delta \bar{h}=0$) which are the
absolute minima of the energy for fixed $Q$ \cite{s-coleman}. For
$\omega=\omega_{0}$
the effective potential $V_{\omega}$ has two degenerate minima one at the origin
($\bar{S}=0,\bar{h}=0$), and the other at  $(\bar{S}=S_{0}\neq0, \bar{h}=h_{0}\neq
0)$. To determine $S_{0}$ and $h_{0}$ it is convenient to introduce a polar
coordinate system in two-dimensional field space by defining $\tan\theta\equiv
\bar{h}/\bar{S}$ and $H\equiv\sqrt{\bar{S}^{2}+\bar{h}^{2}}$ \cite{mssm-Q}.
Expressing $\mu^{2}$ (10) in
terms of $H$ and $\theta$ one obtains 
\begin{eqnarray}
\mu^{2}(H)=\frac{1}{\cos^{2}\theta}\{m_{H}^{2}+A_{H}H+
\frac{1}{2}\lambda_{H}H^{2}\}
\end{eqnarray}
where
\begin{eqnarray}
m_{H}^{2}&=&(m_{s}^{2}-\kappa
v^{2})\cos^{2}\theta+m_{h}^{2}\sin^{2}\theta\nonumber\\
A_{H}&=&(\lambda \sin^{2}\theta-\kappa\cos^{2}\theta)v\sin\theta\\
\lambda_{H}&=&\lambda_{s}\cos^{4}\theta+\lambda \sin^{4}\theta -2\kappa
\sin^{2}\theta \cos^{2}\theta\nonumber\;.
\end{eqnarray}
It is easy to see that $\mu^{2}$ is minimized for $H=H_{0}\equiv
-2A_{H}/\lambda_{H}$ which remains non-vanishing as long as $A_{H}\neq 0$. Moreover
it is always positive if $\kappa$ is sufficiently large, $\kappa\simgt \lambda
\tan^{2}\theta$. This proves that there is a field configuration ($S_{0},h_{0}$)
away from the origin and minimizing the quantity $\mu^{2}$ (12). Correspondingly,
one has 
\begin{eqnarray}
\omega_{0}=\frac{1}{\cos\theta}(m_{H}^{2}-\frac{2 A_{H}^{2}}{\lambda_{H}})^{1/2}
\end{eqnarray}
which is always real as long as $U(1)_{s}$ is an exact symmetry of the
true electroweak vacuum, that is, $A_{H}^{2}< 4 \lambda_{H} m_{H}^{2}/9$. For
$\omega=\omega_{0}$, the effective potential has two minima at $H=0$ and $H=H_{0}$
between which it is maximized at $H=H_{0}/2$ with a value $A_{H}^{4}/4
\lambda_{H}^{3}$. As $\omega$ takes values larger than $\omega_{0}$, the minimum at
$H=H_{0}$ becomes the negative global minimum of the potential. The transition
of the system form the original minimum $H=0$ to the new global minimum proceeds
through the quantum tunneling. The corresponding equations of motion can be
identified with those of a bounce \cite{bounce}. As long as $\omega_{0}\leq \omega
\leq \sqrt{V''(H=0)}$
there is always a bounce solution \cite{small-Q} having spherical symmetry
$H(\vec{x})=H(r)\;, r=\sqrt{\vec{x}^{2}}$ \cite{s-coleman2} and a localized nature
$H(r)\rightarrow 0$ and $r\rightarrow\infty$. The resulting object is a lump of $H$
matter with finite mass $M(Q)=\hat{\mu}(Q)Q$ where $\hat{\mu}(Q)<m_{H}$ always.
Moreover, $\hat{\mu}\rightarrow \mu(S_{0},h_{0})$ for $Q\rightarrow \infty$
\cite{s-coleman}, and has a $Q$-dependent expression for smaller values of $Q$
\cite{small-Q}. Irrespective of the detailed expressions for its extensive
parameters, the Q-ball in question is a spherically symmetric object with a finite
spatial extension characterized by its radius $R(Q)$, and a finite mass $M(Q)$
\cite{bounce,s-coleman,small-Q,mssm-Q}.

The effective potential $V_{\omega_{0}}$ depends only on
$\lambda_{H}$ and $A_{H}$, in particular, it is independent of $m_{H}^{2}$. 
This $m_{H}^{2}$ independence of $V_{\omega_{0}}$ leaves
$m_{s}^{2}$ largely free as it does not affect the possible transitions between the
two degenerate minima. For the calculation of the Higgs invisible decay rate
$\Gamma(h\rightarrow S S)$ in the true electroweak vacuum it is necessary to have
$m_{h}^{2}\geq 4(m_{s}^{2}-\kappa v^{2})$. However, formation of the electroweak
Q-ball does not have such a strong requirement, it requires only the right hand side
of this inequality be positive in order to leave $U(1)_{s}$ unbroken. Unlike the
effective
potential $V_{\omega_{0}}$, however, $\omega_{0}$ (14) has an explicit dependence
on $m_{s}^{2}$. The physically relevant interval for $\omega$ is $\omega_{0}\leq
\omega \leq m_{H}$ \cite{small-Q}, and mass per unit charge $\hat{\mu}(Q)$
interpolates  between the two extremes of $\omega$. For
$m_{s}^{2}>>v^{2}$, the allowed interval for $\omega$ shrinks to a narrow
interval just below $m_{H}$.
In this large $m_{s}^{2}$ limit the total charge $Q$ necessarily becomes 
vanishingly small in which case the semiclassical analysis presented above can be
invalidated.
To avoid such an extreme (and also useless) case we assume that $m_{s}^{2}$ and
$v^{2}$ are not at
diversely different mass scales.

It is the non-vanishing of the trilinear coupling $A_{H}$ that guarantees the
existence of Q-balls in the electroweak vacuum. As is seen from (13) $A_{H}$ is
proportional to $\sin\theta$ and $v$. Vanishing of $\sin\theta$ means the neglect of
the
Higgs field in constructing the Q-ball solution. Similarly, vanishing of $v$ means
the absence of the electroweak phase transition. In both cases the electroweak
Lagrangian (3) cannot accomodate a Q-ball solution. Hence, the existence of the
Q-ball
type solitons in the Lagrangian (3) is triggered essentially by the electroweak
phase transition so we call them electroweak Q-balls from now on. 

The analysis above proves that the electroweak Q-balls exist. To extract
the necessary information
about their relevance to the real world one should investigate its interactions with
the environment. First we discuss the stability of the electroweak Q-ball. By
construction, Q-ball is the state of minimal energy in the scalar sector of the
theory. Hence, by energy conservation, it cannot decay to bosons, in particular, its
constitutents. More importantly, any scalar produced in the true vacuum by some
scattering process rapidly escapes to the Q-ball as a statement of the Bose
statistics. Importance of this statement for the Higgs phenomenology is that, even
if the SM Higgs particle is produced by some future collider it can immediately
escape to the Q-ball on its way to the detector. This observation comes by no
surprise because such models were proposed to account for the non-observation 
of Higgs at collider searches by its large invisible decay rate
\cite{old-hiding,new-hiding}. After making  these observations about the bosonic
sector of
the SM, it remains to discuss fermionic instability channels. First of all, one
notes that there is no fermion (and also boson) in the SM particle spectrum which
has nonzero $U(1)_{s}$ charge. This proves that there is no decay mode which can
cause an erosion of the charge contained in the Q-matter. Since the charge remains
unchanged always, one concludes that Q-ball is absolutely stable \cite{s-coleman}.
The electroweak Q-ball would evaporate if $U(1)_{s}$ symmetry were identified with
$U(1)_{B,L}$ as in the supersymmetric theories \cite{mssm-Q} or other models
\cite{evapo}. This very stability of the electroweak Q-ball implies that it can
survive without dispersion for rather long time intervals at the cosmic scales.

To extract further information about the physical properties of the electroweak
Q-ball it may be convenient to study the scattering of the observable particles from
its bulk. One notes that the Q-matter in the electroweak Q-ball has two components;
the SM
Higgs $\bar{h}$ and the singlet $\bar{S}$ both having nonzero values over the
spatial extension
of the soliton. While the singlet provides the absolute stability of the electroweak
Q-ball, the Higgs component is responsible for communication with the observable
sector of the SM, that is, fermions and gauge bosons. The remarkable thing about the
electroweak Q-ball is that any observable particle incident on it gets reflected
through its interactions only with the Higgs boson, without feeling the presence of
the
singlet. The coupling between the SM particles and Higgs is known for every species
\cite{hunting}: all fermions generically couple as $g m_{f}/2 M_{W}$ and the massive 
vector 
boson $W$ ($Z$) as $g M_{W}$ $(g M_{Z}/\cos\theta_{W})$. Obviously photon and gluon
do not have tree-level couplings due to the electric neutrality and colorlessness
of the Q-ball. Therefore, massless fermions, photon and gluon can
have couplings only at the loop level through effective $h\bar{f}f$, $h\gamma\gamma$
and $h g g$ vertices \cite{hunting}. Appearently, during all these scattering events
momentum conservation is provided by the emission of sound waves from the
electroweak Q-ball \cite{s-coleman}. Due to the asymptotic freedom, quarks and gluon
cannot have isolated free-particle states and their coupling strengths to Higgs
boson are
relevant only for studying the scatterings of hadrons from the Q-ball.
The gross observation about the interaction between electroweak Q-ball and the
SM particle spectrum is that a typical scattering process 
\begin{eqnarray}
\mbox{particle}+\mbox{Q-ball}\rightarrow \mbox{particle}+\mbox{Q-ball+sound waves}
\end{eqnarray}
proceeds essentially with the weak interactions because electromagnetic and strong
interactions can arise only at the loop level. Hence, the electoweak Q-ball has
essentially weak and gravitational interactions, and from the view point of   
the SM particle spectrum, it is a weakly interacting stable lump of non-baryonic matter.
These
properties of the electroweak Q-ball reminds at once one of the missing mass in the
universe, that is, the dark matter.

There is strong evidence from a variety of sources for a large amount of dark matter
in the universe \cite{dark1}. There is also extensive evidence that a substantial
amount of the dark matter is non-baryonic. Models of galaxy formation classify the
non-baryonic dark matter as hot and cold depending on if the constitutents have
relativistic or non-relativistic velocities, respectively. This classification
can be reduced to the language of masses of the dark matter particles with a
dividing line $m_{DM}\sim 1\;keV$.
If the dark matter particles are their own anti-particles,
and they are in thermal equilibrium with the radiation then their relic abundance is
determined mainly by their annihilation cross section. The value of the annihilation
cross section needed to make the relic abundance of the dark matter close to unity
is remarkably close to one would expect for a weakly interacting massive particle
(WIMP) with a mass $m_{DM}\sim M_{Z}$. The two best known and most studied cold dark
matter candidates are neutralino \cite{neut} and axion \cite{axs} both qualifying to
be WIMP's. If R-parity is conserved the neutralino with a mass in hundreds of GeV is
a WIMP candidate. Similarly, axion arises in extensions of the SM to solve strong
CP problem, and it has a rather small mass of the order of $10^{-5}\; \mbox{eV}$.
Recently, L- balls occuring in the scalar sectors of the supersymmetric theories are
idenfied as dark matter candidates \cite{dark-Q,enqvist2}. If the scalar potential
is flat
the L-ball can be large enough to survive until the present time in spite of
the evaporation to light leptons. Recalling the properties of the electroweak Q-ball
one
observes that it behaves as a collection of some $Q$ WIMP's each with mass
$M(Q)/Q=\hat{\mu}(Q)$. For consistency one needs $\hat{\mu}(Q)\sim M_{Z}$, which
establishes another requirement for having $m_{s}^{2}$ at the weak scale. Unlike the
L-balls of the supersymmetric models, the electroweak Q-ball does not suffer from
evaporation to light fermions so it is a stable dark matter candidate. Guided by the
analysis of \cite{kolb} one would say that the small electroweak Q-balls can be
produced
copiously in the early universe during the electroweak phase trasition, and they
subsequently merge to form big Q-balls \cite{dark-Q}. Due to their absolute 
stability they
can survive until the present time and contribute to the total mass density of the
universe in the form of dark matter \cite{dark-Q,kolb2}.
It is the future
astronomical observations about the dark matter that will specify its luminious,
baryonic and non-baryonic proportions. After having such quantitative
information about the dark matter that it
will be possible to test the predictions of the electroweak Q-ball.
However, at the present precision of the observations, dark matter is essentially
non-luminious and non-baryonic so that the electroweak Q-ball can be a
candidate to explain its existence in the universe. 

As usual proton is absolutely stable because there is no $U(1)_{L,B}$ violating  
interactions in the SM Lagrangian. However, in the presence of the SM singlet
one can introduce a non-renormalizable interaction of the form
\begin{eqnarray}
\Delta {\cal{L}}=\kappa^{\prime} \frac{|S|^{2}}{M_{X}}\bar{Q}L + h.c.
\end{eqnarray}
where $M_{X}>>m_{s}$ is a large mass scale, $\kappa^{\prime}$ is a Yukawa
coupling, and $Q$ and $L$ are light quark and lepton with $m_{Q}>m_{L}$. $\Delta
{\cal{L}}$ breaks the B- and L- symmetries of the SM Lagrangian but preserves the
$U(1)_{s}$ symmetry. In the true elecetroweak vacuum $Q\rightarrow L$ transition
occurs only at the two loop level and its rate is small \cite{dvali}. However,
inside the electroweak Q-ball $S$ has a non-vanishing VEV and $Q\rightarrow L$ type
transitions can occur at the tree level as illustrated by (15). More interestingly,
nucleon scattering from the Q-ball can realize $Q\rightarrow L$ transition with a 
probability $\sim {\kappa^{\prime}}^{2} \frac{\bar{S}^{4}}{m_{Q}^{2}M_{X}^{2}}$
\cite{dvali}. The ultimate existence of such L- and B-violating interactions will be
tested with future experiments on the dark matter.

With the ending of LEP2 period without a signal, search for the Higgs boson will
continue at the LHC \cite{lhc}. In near future, the LHC will be searching for
the Higgs
signal in the mass range $M_{Z}\simlt m_{h}\simlt 2 M_{Z}$ expecting to observe the
Higgs resonance in gluon-gluon fusion to photon pairs. In this search strategy
observation of the Higgs resonance is essential to extract the Higgs signal from the
large irreducible background. In future, if the LHC fails to find a Higgs signal,
possibility of a weak-scale hidden sector will be strengthened 
\cite{old-hiding,new-hiding,ghinculov}. Besides the Higgs discovery potential of
the LHC, there are
strong theoretical arguments stating that a linear collider working at the
center-of-mass energy $\sqrt{s}=500\;\mbox{GeV}$ with an integrated annual 
limunosity of $500\; fb^{-1}$, such as the TESLA collider \cite{tesla}, will
definitely find a signal for the Higgs boson \cite{jose} independent of the
complexity of the Higgs sector and Higgs boson decay modes. Hence, the no-lose
theorem of \cite{jose}, armed with the recently proposed TESLA collider
\cite{tesla}, forms a testing ground for the existence of a perturbative Higgs
sector. If the results of \cite{jose} cannot be confirmed at a future linear
collider then the present status of the symmetry  breaking sector of the SM will be
questionable, and a possibility of having a hidden electroweak sector will increase.

In this work we investigated the implications of a weak-scale hidden sector for dark
matter searches, proton-to-lepton transitions, and Higgs phenomenology at the future
colliders. The main object of the discussion has been the electroweak Q-ball, a
non-topological, extended, absolutely stable object having mainly weak and
gravitational interactions. The formation of the electroweak Q-ball is triggered by
the electroweak phase transition. The indispensable component of the analysis, that
is, the SM singlet has a mass at the electroweak scale, as required by both the
Q-ball formation and dark matter phenomenology. If the future collider search fails
to find an observable signal for the Higgs boson, possibility of a weak-scale
hidden sector, like the one discussed above,  will increase. As discussed in the
text, supersymmetric theories generically and naturally accomodate Q-balls
corresponding to the exact global symmetries, $U(1)_{B,L}$, of the low-energy
theory. However, it is known that these Q-balls necessarily evaporate by emitting
lepton or baryon number from their surface. Thus, despite the existence of
Higgs-sfermion trilinear couplings in the Lagrangian, only in the case of flat
potentials that one can construct big enough Q-balls that can survive until the
present time after being formed in the early universe. Unlike the $U(1)_{B,L}$
symmetries leading naturally to Q-balls in supersymmetric theories, the electroweak
Q-ball is constructed by postulating an extra $U(1)$ whose origin is unknown.
However, it is with this $U(1)_{s}$ group that the electroweak Q-ball does not
suffer from evaporation, and is an absolutely stable object [$U(1)_s$ has no
relation to $U(1)_{B,L}$]. Despite the unknown nature of the singlet and its
couplings, the electroweak Q-ball interacts with the visible matter only through the
Higgs field, and couplings of the SM particle spectrum to Higgs are already known.
Therefore, existence of the electroweak Q-balls can be directly tested with the
future astronomical observations on the dark matter. Indeed, the experimental data
concerning the luminious, baryonic and non-baryonic proportions of the dark matter
should be derivable from the coupling strengths of the observable matter to the Q-ball.
For Q-ball formation, the SM Higgs sector is not the only possiblity, in
fact, a similar analysis can be carried out for two-doublet models in which there
are two CP-even and a CP-odd scalar Higgs bosons which can contribute to the
associated Q-matter. 
\newline
\newline
\newline
Author would like to thank S. Randjbar-Daemi, A. V. Rubakov, G.
Senjanovi$\stackrel{'}{c}$, and A. Smirnov for helpful discussions.

\end{document}